\documentclass[10pt]{article}
\usepackage{latexsym,graphicx}
\newcommand{\be}{\begin{equation}}
\newcommand{\ee}{\end{equation}}
\def\n{\noindent}
\catcode `\@=11
\catcode `\@=12
\begin{document}
\begin{center}
\large{\bf{Magnetized Bianchi Type $VI_{0}$ Barotropic Massive String Universe 
with Decaying Vacuum Energy Density $\Lambda$}} \\
\vspace{10mm}
\normalsize{Anirudh Pradhan $^1$ and Raj Bali $^2$} \\
\vspace{5mm}
\normalsize{$^{1}$Department of Mathematics, Hindu Post-graduate College, 
Zamania-232 331, Ghazipur, India \\
E-mail : pradhan@iucaa.ernet.in} \\
\vspace{5mm}
\normalsize{$^{2}$Department of Mathematics, University of Rajasthan, Jaipur-302 004, India \\
E-mail : balir5@yahoo.co.in}\\
\end{center}
\vspace{10mm}
\begin{abstract}
Bianchi type $VI_{0}$ massive string cosmological models using the technique given by Letelier (1983) 
with magnetic field are investigated. To get the deterministic models, we assume that the expansion 
($\theta$) in the model is proportional to the shear ($\sigma$) and also the fluid obeys the barotropic 
equation of state. It was found that vacuum energy density $\Lambda \propto \frac{1}{t^{2}}$ which matches 
with natural units. The behaviour of the models from physical and geometrical aspects in presence 
and absence of magnetic field is also discussed. 
\end{abstract}
\smallskip
\n Keywords : Massive string, Bianchi type $VI_{0}$ universe, Magnetic field \\
\n PACS number: 98.80.Cq, 04.20.-q 
\section{Introduction}
The problem of the cosmological constant is one of the most salient and unsettled problems in 
cosmology. The smallness of the effective cosmological constant recently observed 
$(\Lambda_0 \leq  10^{-56} {\rm cm^{-2}})$ constitutes the most difficult problems involving 
cosmology and elementary particle physics theory. To explain the striking cancellation between 
the ``bare'' cosmological constant and the ordinary vacuum energy contributions of the quantum 
fields, many mechanisms have been  proposed during last few years \cite{ref1}. The ``cosmological 
constant problem'' can be expressed as the discrepancy between the negligible value of $\Lambda$ 
has for the present universe (as can be seen by the successes of Newton's theory of gravitation 
\cite{ref2}) and the values $10^{50}$ larger expected by the Glashow-Salam-Weinberg model \cite{ref3} 
or by grand unified theory (GUT) where it should be $10^{107}$  larger \cite{ref4}. The cosmological 
term $\Lambda$ is then small at the present epoch. The problem 
in this approach is to determine the right dependence of $\Lambda$ upon $S$ or $t$.  Recent 
observations of Type Ia supernovae (Perlmutter et al. \cite{ref5}, Riess et al. \cite{ref6}) and 
measurements of the cosmic microwave background \cite{ref7} suggest that the universe is 
an accelerating expansion phase \cite{ref8}. 
\newline
\par
Several ans$\ddot{a}$tz have been proposed in which the $\Lambda$ term decays 
with time (see Refs. Gasperini \cite{ref9}, Berman \cite{ref10}$-$\cite{ref12}, 
Berman et al. \cite{ref13}$-$\cite{ref15}, Freese et al. \cite{ref16}, $\ddot{O}$zer and 
Taha \cite{ref17}, Ratra and Peebles \cite{ref18}, Chen and Hu \cite{ref19}, Abdussattar and 
Vishwakarma \cite{ref20}, Gariel and Le Denmat \cite{ref21}, Pradhan et al. \cite{ref22}). Of 
the special interest is the ans$\ddot{a}$tz $\Lambda \propto S^{-2}$ (where $S$ is the scale 
factor of the Robertson-Walker metric) by Chen and Wu \cite{ref19}, which has been 
considered/modified by several authors ( Abdel-Rahaman \cite{ref23}, 
Carvalho et al. \cite{ref24}, Silveira and Waga \cite{ref25},
Vishwakarma \cite{ref26}).  
\newline
\par
One of the outstanding problems in cosmology today is developing a more precise understanding 
of structure formation in the universe, that is, the origin of galaxies and other large-scale 
structures. Existing theories for the structure formation of the Universe fall into two 
categories, based either upon the amplification of quantum fluctuations in a scalar field during 
{\it inflation}, or upon symmetry breaking phase transition in the early Universe which leads to the 
formation of {\it topological defects} such as domain walls, cosmic strings, monopoles, textures and 
other 'hybrid' creatures. Cosmic strings play an important role in the study of the early universe. 
These arise during the phase transition after the big bang explosion as the temperature goes down 
below some critical temperature as predicted by grand unified theories (see Refs. Zel'dovich 
et al. \cite{ref27}, Kibble \cite{ref28,ref29}, Everett \cite{ref30}, Vilenkin \cite{ref31}). 
It is believed that cosmic strings give rise to density perturbations which lead to formation of 
galaxies (Zel'dovich \cite{ref32}). These cosmic strings have stress energy and couple to the 
gravitational field. Therefore, it is interesting to study the gravitational effect 
which arises from strings. The general treatment of strings was initiated by Letelier \cite{ref33,ref34} 
and Stachel \cite{ref35}. 
\newline
\par
The occurrence of magnetic fields on galactic scale is well-established fact today, and their 
importance for a variety of astrophysical phenomena is generally acknowledged. Several authors 
(Zeldovich \cite{ref36}, Harrison \cite{ref37}, Misner, Thorne and Wheeler \cite{ref38}, 
Asseo and Sol \cite{ref39}, Pudritz and Silk \cite{ref40}, Kim, Tribble, and Kronberg \cite{ref41}, 
Perley, and Taylor \cite{ref42}, Kronberg, Perry, and Zukowski \cite{ref43}, Wolfe, Lanzetta and 
Oren \cite{ref44}, Kulsrud, Cen, Ostriker and Ryu \cite{ref45} and Barrow \cite{ref46}) have pointed 
out the importance of magnetic field in different context. As a natural consequences, we should include 
magnetic fields in the energy-momentum tensor of the early universe. The string cosmological models with 
a magnetic field are also discussed by Benerjee et al. \cite{ref47}, Chakraborty \cite{ref48}, Tikekar 
and Patel (\cite{ref49,ref50}, Patel and Maharaj \cite{ref51} Singh and Singh \cite{ref52}. 
\newline
\par
Recently, Bali et al. \cite{ref53}$-$\cite{ref57}, Pradhan et al. \cite{ref58,ref59}, 
Yadav et al. \cite{ref60} and Pradhan \cite{ref61} have investigated Bianchi type I, II, III, V,IX  
and cylindrically symmetric magnetized string cosmological models in presence and absence of 
magnetic field. Tikekar and Patel \cite{ref50} have investigated some solutions for Bianchi 
type $VI_{0}$ cosmology in presence and absence of magnetic field. In this paper we have derived 
some Bianchi type $VI_{0}$ string cosmological models for perfect fluid distribution in presence and 
absence of magnetic field and discussed the variation of $\Lambda$ with time. This paper is organized 
as follows: The metric and field equations are presented in Section $2$. In Section $3$, we deal with 
the solution of the field equations in presence of magnetic field. In Section $4$, we have described 
some geometric and physical behaviour of the model. Section $5$ includes the solution in absence of 
magnetic field. In Section $6$, we have discussed the variation of cosmological constant $\Lambda$ 
with time in presence and absence of magnetic field. In the last Section $7$, concluding remarks are given.  
\section{The Metric and Field  Equations}
We consider the Bianchi Type $VI_{0}$ metric in the form 
\begin{equation}
\label{eq1}
ds^{2} =  - dt^{2} + A^{2}(t) dx^{2} + B^{2}(t) e^{2x} dy^{2} + C^{2}(t) e^{-2x} dz^{2}.
\end{equation}
The energy-momentum tensor for a cloud of strings in presence of magnetic field 
is taken into the form 
\begin{equation}
\label{eq2}
T_{ik} = (\rho + p)v_{i}v_{k} + p g_{ik} - \lambda x_{i}x_{k} + 
v_{i} v_{k}) +  [g^{lm} F_{il} F_{km} - \frac{1}{4} g_{ik} F_{lm} F^{lm}],
\end{equation}
where $v_{i}$ and $x_{i}$ satisfy conditions
\begin{equation}
\label{eq3}
v^{i} v_{i} = - x^{i} x_{i} = -1, \, \, \, v^{i} x_{i} = 0.
\end{equation}
In equations (\ref{eq2}), $p$ is isotropic pressure, $\rho$ is rest energy density for a cloud 
strings, $\lambda$ is the string tension density, $F_{ij}$ is the electromagnetic field tensor, 
$x^{i}$ is a unit space-like vector representing the direction of string, and $v^{i}$ is the 
four velocity vector satisfying the relation
\begin{equation}
\label{eq4}
g_{ij} v^{i} v^{j} = -1.
\end{equation}
Here, the co-moving coordinates are taken to be $v^{1} = 0 = v^{2} = v^{3}$ and $v^{4} = 1$ and 
$x^{i} = (\frac{1}{A}, 0, 0, 0)$. 
The Maxwell's equations  
\begin{equation}
\label{eq5}
F_{ij;k} + F_{jk; i} + F_{ki;j} = 0,
\end{equation}
\begin{equation}
\label{eq6}
F^{ik}_{;k} = 0,
\end{equation}
are satisfied by
\begin{equation}
\label{eq7}
F_{23} = K \mbox{(say)} = \mbox {constant}, 
\end{equation}
where a semicolon ($;$) stands for covariant differentiation. \\

The Einstein's field equations (with $\frac{8\pi G}{c^{4}} = 1$) 
\begin{equation}
\label{eq8}
R^{j}_{i} - \frac{1}{2} R g^{j}_{i} = - T^{j}_{i} - \Lambda g^{j}_{i},
\end{equation}
for the line-element(\ref{eq1}) lead to the following system of equations:  
\begin{equation}
\label{eq9}
\frac{B_{44}}{B} + \frac{C_{44}}{C} + \frac{B_{4} C_{4}}{B C} + \frac{1}{A^{2}} = - 
\left[p - \lambda - \frac{K^{2}}{2 B^{2} C^{2}}\right] - \Lambda,
\end{equation}
\begin{equation}
\label{eq10}
\frac{A_{44}}{A} + \frac{C_{44}}{C} + \frac{A_{4} C_{4}}{A C}  - \frac{1}{A^{2}} =  -
\left[p + \frac{K^{2}}{2 B^{2} C^{2}}\right] - \Lambda, 
\end{equation}
\begin{equation}
\label{eq11}
\frac{A_{44}}{A} + \frac{B_{44}}{B} + \frac{A_{4} B_{4}}{A B} - \frac{1}{A^{2}}=  - 
\left[p + \frac{K^{2}}{2 B^{2} C^{2}}\right] - \Lambda,
\end{equation}
\begin{equation}
\label{eq12}
\frac{A_{4}B_{4}}{A B} + \frac{B_{4} C_{4}}{B C} + \frac{C_{4} A_{4}}{C A} - \frac{1}{A^{2}} = 
 \left[\rho + \frac{K^{2}}{2 B^{2} C^{2}}\right] - \Lambda,
\end{equation}
\begin{equation}
\label{eq13}
\frac{1}{A}\left[\frac{C_{4}}{C} - \frac{B_{4}}{B}\right] = 0,
\end{equation}
where the sub indice $4$ in A, B, C denotes ordinary differentiation with respect to $t$.
The velocity field $v^{i}$ is irrotational. The scalar expansion $\theta$ and components 
of shear $\sigma_{ij}$ are given by
\begin{equation}
\label{eq14}
\theta = \frac{A_{4}}{A} + \frac{B_{4}}{B} + \frac{C_{4}}{C},
\end{equation}
\begin{equation}
\label{eq15}
\sigma_{11} = \frac{A^{2}}{3}\left[\frac{2A_{4}}{A} - \frac{B_{4}}{B} - \frac{C_{4}}{C}\right],
\end{equation}
\begin{equation}
\label{eq16}
\sigma_{22} = \frac{B^{2}}{3}\left[\frac{2B_{4}}{B} - \frac{A_{4}}{A} - \frac{C_{4}}{C}
\right],
\end{equation}
\begin{equation}
\label{eq17}
\sigma_{33} = \frac{C^{2}}{3}\left[\frac{2C_{4}}{C} - \frac{A_{4}}{A} - \frac{B_{4}}{B}\right],
\end{equation}
\begin{equation}
\label{eq18}
\sigma_{44} = 0.
\end{equation}
Therefore
\[
\sigma^{2} = \frac{1}{2}\Biggl[(\sigma^{1}_{~~ 1})^{2} + (\sigma^{2}_{~~ 2})^{2} + (\sigma^{3}_{~~ 3})^{2} + 
(\sigma^{4}_{~~ 4})^{2}\Biggr], 
\]
which leads to
\[
\sigma^{2} = \frac{1}{3}\left[\frac{A_{4}^{2}}{A^{2}} + \frac{B_{4}^{2}}{B^{2}} + 
\frac{C_{4}^{2}}{C^{2}} - \frac{A_{4}B_{4}}{AB} - \frac{B_{4}C_{4}}{BC} - 
\frac{C_{4}A_{4}}{CA}\right]. 
\]
Above relation after using (\ref{eq13}) reduces to
\begin{equation}
\label{eq19}
\sigma^{2} = \frac{1}{\sqrt{3}}\left(\frac{A_{4}}{A} - \frac{B_{4}}{B}\right). 
\end{equation}
\section{Solutions of the Field  Equations}
The field equations (\ref{eq9})-(\ref{eq13}) are a system of five equations with seven 
unknown parameters $A$, $B$, $C$, $\rho$, $p$, $\lambda$ and $\Lambda$. We need 
two additional conditions to obtain explicit solutions of the system. 

Equation (\ref{eq13}) leads to
\begin{equation}
\label{eq20}
C = m B,
\end{equation}
where $m$ is an integrating constant. \\

We first assume that the expansion ($\theta$) is proportional to shear ($\sigma$). This condition and 
Eq. (\ref{eq20}) lead to 
\begin{equation}
\label{eq21}
\frac{1}{\sqrt{3}}\left(\frac{A_{4}}{A} - \frac{B_{4}}{B}\right) = \l \left(\frac{A_{4}}{A} 
+ \frac{2B_{4}}{B}\right)
\end{equation}
which yields to
\begin{equation}
\label{eq22}
\frac{A_{4}}{A} = n \frac{B_{4}}{B},
\end{equation}
where $n = \frac{(2\l \sqrt{3} + 1)}{(1 - \l \sqrt{3})}$ and $\l$ are constants. 
Eq. (\ref{eq22}), after integration,  reduces to
\begin{equation}
\label{eq23}
A = \beta B^{n},
\end{equation}
where $\beta$ is a constant of integration. Eqs. (\ref{eq10}) and (\ref{eq12}) lead to
\begin{equation}
\label{eq24}
p = - \frac{K^{2}}{2 B^{2} C^{2}} - \left(\frac{A_{44}}{A} + \frac{C_{44}}{C} + 
\frac{A_{4} C_{4}}{A C}  - \frac{1}{A^{2}}\right) - \Lambda,
\end{equation}
and
\begin{equation}
\label{eq25}
\rho = \frac{A_{4}B_{4}}{A B} + \frac{B_{4} C_{4}}{B C} + \frac{C_{4} A_{4}}{C A} - 
\frac{1}{A^{2}} - \frac{K^{2}}{2 B^{2} C^{2}} + \Lambda,
\end{equation}
respectively. Now let us consider that the fluid obeys the barotropic equation of state 
\begin{equation}
\label{eq26}
p = \gamma \rho,
\end{equation}
where $\gamma (\gamma \leq 0 \leq 1)$ is a constant. Eqs. (\ref{eq24}) to (\ref{eq26}) lead to 

\[
\frac{A_{44}}{A} + \frac{C_{44}}{C} + (1 + \gamma)\frac{A_{4}C_{4}}{AC} + \gamma\left(\frac{A_{4}B_{4}}
{AB} + \frac{B_{4}C_{4}}{BC}\right) - (1 + \gamma)\frac{1}{A^{2}} + 
\] 
\begin{equation}
\label{eq27}
(1 - \gamma)\frac{K^{2}}
{2B^{2}C^{2}} + (1 + \gamma)\Lambda = 0.
\end{equation}
Eq. (\ref{eq27}) with the help of (\ref{eq20}) and (\ref{eq23}) reduces to   
\begin{equation}
\label{eq28}
2B_{44} + \frac{2(n^{2} + 2\gamma n + \gamma)}{(n + 1)}\frac{B_{4}^{2}}{B^{2}} = 
\frac{2(1 + \gamma)}{\beta^{2} B^{2n - 1}} + \frac{(1 - \gamma)K^{2}}{m^{2}B^{3}} 
+ 2\l_{0} B,
\end{equation}
where $\l_{0} = (1 + \gamma)\Lambda $. \\
Let us consider $B_{4} = f(B)$ and $f' = \frac {df}{dB}$. Hence Eq. (\ref{eq28}) takes the form
\begin{equation}
\label{eq29}
\frac{d}{df}(f^{2}) + \frac{2\alpha}{B} f^{2} = \frac{2(1 + \gamma)}{\beta^{2}B^{2n - 1}} 
+ \frac{(1 - \gamma)K^{2}}{m^{2}B^{3}} + 2\l_{0} B,
\end{equation}
where $ \alpha = \frac{(n^{2} + 2n \gamma + \gamma)}{(n + 1)}$. Eq. (\ref{eq29}) after integrating 
reduces to
\begin{equation}
\label{eq30}
f^{2} = \frac{2(1 + \gamma)B^{-2n + 2}}{\beta^{2}(2\alpha - 2n + 2)} + \frac{(1 - \gamma)K^{2}}
{2m^{2}(\alpha - 1)} + \frac{\l_{0}B^{2}}{(\alpha + 1)} + M B^{-2\alpha},~ ~ \gamma \ne 1, 
\end{equation}
where $M$ is an integrating constant. To get deterministic solution in terms of cosmic string $t$, 
we suppose $M = 0$ without any loss of generality. In this case Eq. 
(\ref{eq30}) takes the form
\begin{equation}
\label{eq31}
f^{2} = a B^{-2(n - 1)} + b B^{-2} + k B^{2},
\end{equation}
where
$$ a = \frac{2(1 + \gamma)}{\beta^{2}(2\alpha - 2n + 2)}, ~ ~  b = \frac{(1 - \gamma)K^{2}}
{2m^{2}(\alpha - 1)}, ~ ~ k = \frac{(1 + \gamma)\Lambda }{(\alpha + 1)}.$$
Therefore, we have
\begin{equation}
\label{eq32}
\frac{dB}{\sqrt{a B^{-2(n - 1)} + b B^{-2} + k B^{2}}} = dt.
\end{equation}  
To get deterministic solution, we assume $n = 2$. In this case integrating Eq. (\ref{eq32}), we obtain
\begin{equation}
\label{eq33}
B^{2} = \sqrt{(a + b)} \frac{\sinh{(2\sqrt{k} t)}}{\sqrt{k}}.
\end{equation}
Hence, we have
\begin{equation}
\label{eq34}
C^{2} = m^{2} \sqrt{(a + b)} \frac{\sinh{(2\sqrt{k} t)}}{\sqrt{k}},
\end{equation}
\begin{equation}
\label{eq35}
A^{2} = \beta^{2} (a + b) \frac{\sinh^{2}{(2\sqrt{k} t)}}{k},
\end{equation}
where $k > 0$ without any loss of generality. \\

Therefore, the metric (\ref{eq1}), in presence of magnetic field, reduces to the form
\[
ds^{2} = - dt^{2} + \beta^{2} (a + b) ~ \frac{\sinh^{2}{(2\sqrt{k} t)}}{k} ~ dx^{2} + 
\]
\begin{equation}
\label{eq36}
\sqrt{(a + b)} ~ \frac{\sinh{(2\sqrt{k} t)}}{\sqrt{k}} ~ e^{2x} ~ dy^{2} + m^{2} \sqrt{(a + b)} ~ 
\frac{\sinh{(2\sqrt{k} t)}}{\sqrt{k}} ~ e^{-2x} ~ dz^{2}.
\end{equation}
\section{The Geometric and Physical Significance of Model}
The pressure ($p$), energy density ($\rho$), the string tension density ($\lambda$), the particle 
density ($\rho_{p}$), the scalar of expansion ($\theta$), the shear tensor ($\sigma$) and the 
proper volume ($V^{3}$) for the model (\ref{eq36}) are given by  
\[
p = \left[\frac{k}{\beta^{2}(a + b)} - \frac{K^{2}k}{2m^{2}(a + b)}\right] \coth^{2}{(2\sqrt{k} t)} +
\] 
\begin{equation}
\label{eq37}
\left[\frac{K^{2}}{2m(a + b)} - \frac{1}{\beta^{2}(a + b)} - 8\right]k - \Lambda,
\end{equation}
\[
\rho = \Biggl[5k - \frac{k}{(a + b)}\left(\frac{K^{2}}{2m^{2}} + \frac{1}{\beta^{2}}\right)\Biggr]
\coth^{2}{(2\sqrt{k} t)} + 
\]
\begin{equation}
\label{eq38}
\frac{k}{(a + b)}\left(\frac{K^{2}}{2m^{2}} + \frac{1}{\beta^{2}}\right) 
+ \Lambda,
\end{equation}
where $p = \gamma \rho$ is satisfied by (\ref{eq27}).
\[
\lambda = \left[\frac{2k}{\beta^{2}(a + b)} - \frac{K^{2}k}{m^{2}(a + b)} - k\right] 
\coth^{2}{(2\sqrt{k} t)} + 
\]
\begin{equation}
\label{eq39}
\left\{\frac{K^{2}k}{m^{2}(a + b)} - \frac{2k}{\beta^{2}(a + b)} - 4k 
\right\},
\end{equation}
\[
\rho_{p} = \rho - \lambda = \left[\frac{K^{2}k}{2m^{2}(a + b)} - \frac{3k}{\beta^{2}(a + b)} + k\right] 
\coth^{2}{(2\sqrt{k} t)}  
\]
\begin{equation}
\label{eq40}
+ 9k + \left\{\frac{3k}{\beta^{2}(a + b)} - \frac{K^{2}}{2m^{2}(a + b)}\right\},
\end{equation}
\begin{equation}
\label{eq41}
\theta = 4\sqrt{k}\coth{(2\sqrt{k} t)},
\end{equation}
\begin{equation}
\label{eq42}
\sigma = \sqrt{\frac{k}{3}} \coth{(2\sqrt{k} t)},
\end{equation}
\begin{equation}
\label{eq43}
V^{3} = \frac{\beta m (a + b)}{k}\sinh^{2}{(2\sqrt{k} t)}.
\end{equation}
From Eqs. (\ref{eq30}) and (\ref{eq31}), we obtain
\begin{equation}
\label{eq44}
\frac{\sigma}{\theta} = \mbox{constant}.
\end{equation}
The deceleration parameter is given by
\begin{equation}
\label{eq45}
q = - \frac{\ddot{R}/R}{\dot{R}^{2}/R^{2}} = - \Biggl[\frac{\frac{8k}{3} - \frac{8k}{9}\coth^{2}
{(2\sqrt{k} t)}}{\frac{16k}{9} \coth^{2}{(2\sqrt{k}t)}}\Biggr].
\end{equation}
From (\ref{eq45}), we observe that
$$ q < 0 ~ ~ if ~ ~ \coth^{2}{(2\sqrt{k}t)} < 3 $$
and
$$ q > 0 ~ ~ if ~ ~ \coth^{2}{(2\sqrt{k}t)} > 3 .$$
From (\ref{eq38}), $\rho \geq 0$ implies that
\begin{equation}
\label{eq46}
\coth^{2}{(2\sqrt{k} t)} \leq \Biggl[\frac{\frac{k}{(a + b)}\left(\frac{K^{2}}{2m^{2}} + \frac{1}{\beta^{2}}
\right) + \Lambda}{\frac{k}{(a + b)}\left(\frac{K^{2}}{2m^{2}} + \frac{1}{\beta^{2}}\right) - 5k}\Biggr].
\end{equation}
Also from (\ref{eq40}), $\rho_{p} \geq 0$ implies that
\begin{equation}
\label{eq47} 
\coth^{2}{(2\sqrt{k} t)} \leq \Biggl[\frac{\frac{3k}{\beta^{2}(a + b)} - \frac{K^{2}}{2m^{2}(a + b)} + 9k}
{\frac{3k}{\beta^{2}(a + b)} - \frac{K^{2} k}{2m^{2}(a + b)} - k}\Biggr].
\end{equation}
Thus the energy conditions $\rho \geq 0$, $\rho_{p} \geq 0$ are satisfied under conditions given by 
(\ref{eq46}) and (\ref{eq47}). 

The model (\ref{eq36}) starts with a big bang at $t = 0$. The expansion in the model decreases as 
time increases. The proper volume of the model increases as time increases. Since $\frac{\sigma}{\theta}$ 
= constant, hence the model does not approach isotropy. Since $\rho$, $\lambda$, $\theta$, $\sigma$ tend 
to infinity and $V^{3} \to 0$ at initial epoch $t = 0$, therefore, the model (\ref{eq36}) for massive string 
in presence of magnetic field has Line-singularity (Banerjee et al. \cite{ref47}). For the condition 
$\coth^{2}{(2\sqrt{k}t)} < 3$, the solution gives accelerating model of the universe. It can be easily 
seen that when $\coth^{2}{(2\sqrt{k}t)} > 3$, our solution represents decelerating model of the universe.
\section{Solutions in Absence of Magnetic Field}
In absence of magnetic field, i.e. when $b \to 0$ i.e. $K \to 0$, we obtain
\[
B^{2} = 2\sqrt{2} ~ \frac{\sinh{(2\sqrt{k} t)}}{2\sqrt{k}},
\]
\[
C^{2} = 2m^{2} \sqrt{a} ~ \frac{\sinh{(2\sqrt{k} t)}}{2\sqrt{k}},
\]
\begin{equation}
\label{eq48}
A^{2} = 4a\beta^{2} ~ \frac{\sinh^{2}{(2\sqrt{k} t)}}{4k}.
\end{equation}
Hence, in this case, the geometry of the universe (\ref{eq36}) reduces to
\[
ds^{2} = - dt^{2} + 4 \beta^{2} a  ~ \frac{\sinh^{2}{(2\sqrt{k} t)}}{4k} ~ dx^{2} + 
\]
\begin{equation}
\label{eq49}
2\sqrt{2} ~ \frac{\sinh{(2\sqrt{k} t)}}{2\sqrt{k}} ~ e^{2x} ~ dy^{2} + 2m^{2} \sqrt{a} ~ 
\frac{\sinh{(2\sqrt{k} t)}}{2\sqrt{k}} ~ e^{-2x} ~ dz^{2}.
\end{equation}
The pressure ($p$), energy density ($\rho$), the string tension density ($\lambda$), the particle 
density ($\rho_{p}$), the scalar of expansion ($\theta$), the shear tensor ($\sigma$) and the 
proper volume ($V^{3}$) for the model (\ref{eq49}) are given by  
\begin{equation}
\label{eq50}
p = \frac{k}{a \beta^{2} }\coth^{2}{(2\sqrt{k} t)} - \left(\frac{1}{a \beta^{2}} + 8\right)k - \Lambda,
\end{equation}
\begin{equation}
\label{eq51}
\rho = \left(5k - \frac{k}{a \beta^{2}}\right)\coth^{2}{(2\sqrt{k} t)} + \frac{k}{a \beta^{2}} + \Lambda,
\end{equation}
\begin{equation}
\label{eq52}
\lambda = \left[\frac{2k}{a \beta^{2}} - k \right] 
\coth^{2}{(2\sqrt{k} t)} - \left\{\frac{2k}{a \beta^{2}} + 4k \right\},
\end{equation}
\begin{equation}
\label{eq53}
\rho_{p} = \rho - \lambda = \left[k - \frac{3k}{a \beta^{2}}\right] 
\coth^{2}{(2\sqrt{k} t)} + 9k + \frac{3k}{\beta^{2}a},
\end{equation}
\begin{equation}
\label{eq54}
\theta = 4\sqrt{k}\coth{(2\sqrt{k} t)},
\end{equation}
\begin{equation}
\label{eq55}
\sigma = \sqrt{\frac{k}{3}} \coth{(2\sqrt{k} t)},
\end{equation}
\begin{equation}
\label{eq56}
V^{3} = \frac{\beta m a}{k}\sinh^{2}{(2\sqrt{k} t)}.
\end{equation}
From Eqs. (\ref{eq54}) and (\ref{eq55}), we obtain
\begin{equation}
\label{eq57}
\frac{\sigma}{\theta} = \mbox{constant}.
\end{equation}
From (\ref{eq51}), $\rho \geq 0$ implies that
\begin{equation}
\label{eq58}
\coth^{2}{(2\sqrt{k} t)} \leq \Biggl[\frac{\frac{k}{a \beta^{2}} + \Lambda}
{\frac{k}{a \beta^{2}} - 5k}\Biggr].
\end{equation}
Also from (\ref{eq53}), $\rho_{p} \geq 0$ implies that
\begin{equation}
\label{eq59} 
\coth^{2}{(2\sqrt{k} t)} \leq \Biggl[\frac{\frac{3k}{a \beta^{2}} + a k}
{\frac{3k}{a\beta^{2}} - k}\Biggr]. 
\end{equation}
Thus the energy conditions $\rho \geq 0$, $\rho_{p} \geq 0$ are satisfied under conditions given by 
(\ref{eq58}) and (\ref{eq59}). 

The model (\ref{eq49}) starts with a big bang at $t = 0$ and the expansion in the model decreases as 
time increases. The spatial volume of the model increases as time increases. Since $\frac{\sigma}{\theta}$ 
= constant, hence the anisotropy is maintained throughout. Since $\rho$, $\lambda$, $\theta$, $\sigma$ tend 
to infinity and $V^{3} \to 0$ at initial epoch $t = 0$, therefore, the model (\ref{eq49}) for massive string 
in absence of magnetic field has Line-singularity \cite{ref47}.
\section{Variation of $\Lambda$ with time}
Equations (\ref{eq37}) and (\ref{eq38}) with the use of (\ref{eq26}) reduce to
\begin{equation}
\label{eq60} 
\coth^{2}{(2\sqrt{k} t)} = \left[\frac{\ell + \alpha + 1}{\ell - 5\gamma}\right], 
\end{equation} 
where 
\begin{equation}
\label{eq61} 
\ell = \frac{(1 + \gamma)}{\beta^{2}(a + b)} - \frac{K^{2}(1 - \gamma)}{2m^{2}(a + b)}. 
\end{equation} 
Thus $\ell$ decreases as magnetic field increases. From equation (\ref{eq60}), we obtain
\begin{equation}
\label{eq62} 
2\sqrt{k} t =  \coth^{-1}\left[\frac{\ell + \alpha + 1}{\ell - 5\gamma}\right]^{\frac{1}{2}},
\end{equation} 
where $\ell + \alpha + 1 > \ell - 5\gamma$ implies that $\alpha + 1 + 5\gamma > 0$ which is true.
Putting the value of $k$ in (\ref{eq62}), we obtain
\begin{equation}
\label{eq63} 
\sqrt{\Lambda} t = \frac{\sqrt{1 + \alpha}}{2\sqrt{1 + \gamma}}\coth^{-1}\left[\frac{\ell + \alpha + 1}
{\ell - 5\gamma}\right]^{\frac{1}{2}} =  \mbox{constant},
\end{equation} 
which implies that
\begin{equation}
\label{eq64} 
\Lambda = \frac{L}{t^{2}},
\end{equation} 
where $L$ is constant. Here we observe that when $t \to 0$ then $\Lambda \to \infty$ and when $t \to \infty$ 
then $\Lambda \to 0$. Here $\Lambda \propto \frac{1}{t^{2}}$ which gives fundamental condition supported 
by observations.

In absence of magnetic field i.e. when $ K \to 0$ then 
\begin{equation}
\label{eq65} 
\ell \to \frac{(1 + \gamma)}{\beta^{2}(a + b)} = \mbox{s (say)}. 
\end{equation}
In this case equations (\ref{eq37}) and (\ref{eq38}) with the use of (\ref{eq26}) reduce to
\begin{equation}
\label{eq66} 
\coth^{2}{(2\sqrt{k} t)} = \left[\frac{s + \alpha + 1}{s - 5\gamma}\right], 
\end{equation}
Putting the value of $k$ in (\ref{eq66}), we obtain
\begin{equation}
\label{eq67} 
\sqrt{\Lambda} t = \frac{\sqrt{1 + \alpha}}{2\sqrt{1 + \gamma}}\coth^{-1}\left[\frac{s + \alpha + 1}
{s - 5\gamma}\right]^{\frac{1}{2}} =  \mbox{constant},
\end{equation} 
which implies that
\begin{equation}
\label{eq68} 
\Lambda = \frac{Q}{t^{2}},
\end{equation}
where $Q$ is constant. Here we observe that when $t \to 0$ then $\Lambda \to \infty$ and when $t \to \infty$ 
then $\Lambda \to 0$. Here $\Lambda \propto \frac{1}{t^{2}}$ which gives fundamental condition supported 
by observations. A number of authors have argued in favour of the dependence $\Lambda \to t^{-2}$ in 
different context. It has also be found, by several authors, that when one supposes variable 
gravitational and cosmological ``constant'' in Brans-Dicke theories one finds the relation like equations 
(\ref{eq64}) and (\ref{eq68}). Berman and Som \cite{ref13} pointed out that the relation $\Lambda \to t^{-2}$ 
seems to play a major role in cosmology. In fact, Berman, Som, and Gomide \cite{ref14} found this relation in 
Brans-Dicke static models; Berman \cite{ref10} found it in a static universe with Endo-Fukui modified Brans-Dicke 
cosmology; Berman and Som \cite{ref13} found it again in general Brans-Dicke models which obey the perfect gas 
equation of state \cite{ref11,ref12}. Berman \cite{ref15} also found this relation in general relativity. We have 
derived the same variation of $\Lambda$ with time in massive string cosmology in this article.     
\section{Concluding Remarks}
Some Bianchi type $VI_{0}$ massive string cosmological models with a perfect fluid as the source 
of matter are obtained in presence and absence of magnetic field. Generally, the models 
are expanding, shearing and non-rotating. In presence of perfect fluid it represents an accelerating 
universe during the span of time mentioned below equation (\ref{eq45}) as decelerating factor $q < 0$ and 
it represents decelerating universe as $q > 0$. All the two massive string cosmological models obtained 
in the present study have Line-singularity (Banerjee et al. \cite{ref47}) at the initial epoch $t = 0$. 
The variation of cosmological term in presence and absence of magnetic field is consistent with recent 
observations. To solve the age parameter and density parameter one require the cosmological constant 
to be positive or equivalently the deceleration parameter to be negative. The nature of the cosmological 
constant $\Lambda$ and the energy density $\rho$ have been examined. We have found that the cosmological 
parameter $\Lambda$ varies inversely with the square of time, which matches its natural units. This supports 
the views in favour of the dependence $\Lambda \to t^{-2}$ first expressed by Bertolami \cite{ref62,ref63} 
and later on observed by several authors \cite{ref9}$-$\cite{ref22}. The density is easily adjustable to 
what we observe today, so that there is no need to have recourse to any critical density, and the 
$\Lambda \to t^{-2}$ law guarantees that we may explain why the present value of $\Lambda$ is negligible 
in comparison with the early universe values as required by particle physics.

We have also observed that the magnetic field gives positive contribution to expansion, shear and the free 
gravitational field which die out for large value of $t$ at a slower rate than the corresponding quantities in the 
absence of magnetic field.
\section*{Acknowledgements} 
One of the authors (A. P. ) would like to thank the Harish-Chandra Research Institute, Allahabad, India 
for providing facility where the part of this work was carried out. 

\end{document}